# Interval Reduced-Order Switched Positive Observers for Uncertain Switched Positive Linear Systems

**Naohisa Otsuka, Daiki Kakehi and Przemysław Ignaciuk**

**Abstract:** In this paper, existence conditions and a design procedure of reduced-order switched positive observers for continuous- and discrete-time switched positive linear systems with uncertainty are established. In the analyzed class, arbitrary switching is permitted, whereas the uncertainty expressed via matrix inequalities concerns both the initial state and system parameters. Positive lower and positive upper interval switched observers are obtained. The proposed observers are of $(n - p)$ order, where $n$ is the dimension of the state vector and $p$ is the rank of the output matrix, i.e., $p$-dimensional measurement information. Moreover, as a special case, existence conditions and a design procedure of reduced-order positive observers for uncertain positive linear systems without switching are provided. The theoretical findings are illustrated by two numerical examples for continuous- and discrete-time systems.

## 1. INTRODUCTION

Positive systems are dynamical systems whose state variables take only nonnegative values. They are encountered in various contexts and application fields, e.g., in the study of population dynamics, biology, transportation, chemical, and electrical systems [1–3]. The fundamental results are provided in [4–7]. A particular attention has been given to the observation problem as a key aspect of creating practical positive control systems [8–10]. The positive observation problem is to construct a *positive* observer that will ensure nonnegative state estimates, as dictated by a respective application [11, 12]. In the presence of uncertainty, the positive observation problem for positive linear systems is usually approached by developing *interval* positive observers [13–15]. However, mainly full-order observers are considered [12, 16]. A good example of a corresponding mathematical framework is given in [12], where the existence and design of an interval positive observer that is robust to parameter variations were investigated for the system and output uncertainties modeled by interval-type matrix relations.

In turn, switched systems – an important class of hybrid dynamical systems – comprise a set of subsystems and a rule that orchestrates the structural transitions among them – the switching. The switching behavior can be found in various real-life objects and diverse application areas, e.g., aircraft industry, mobile robot control, and the world of animals [17]. Essential past results on switched systems have been presented in a number of celebrated books [18–25].

The observer design for both continuous- and discrete-time switched linear systems has been studied in [26–30]. The stability of switched positive linear systems has been investigated in [31–37], whereas interval full-order observer design for switched positive linear systems in [38, 39].

However, the models of many physical systems, e.g., in HIV virus therapy, epidemiology [18], network congestion control [40], multi-modal transport [41] are constructed as switched positive linear systems with uncertain parameters. Therefore, from the practical viewpoint, in the state estimation of switched positive systems it is important to consider the following two aspects:

(i) Use the output information in a more efficient way, i.e., to provide an observer only for the states that are not directly read from the output – a reduced-order switched positive observer instead of a full-order positive observer.

(ii) Ensure robustness against parametric uncertainty, i.e., to provide a robust interval positive observer.

With respect to objective (i) only, interval reduced-order switched positive observers for both continuous- and discrete-time switched positive linear systems without uncertainties have been designed in [38, 39]. Meanwhile, separately, condition (ii) has been satisfied by interval full-order switched positive observers, demonstrated robust against uncertainty of subsystem matrices in papers [42, 43]. However, there has been no research on the existence and design of interval switched positive observers that satisfy the above two conditions (i) and (ii), simultaneously.

Naohisa Otsuka is with the Division of Science, School of Schience and Engineering, Tokyo Denki University. Hatoyama-Machi, Hiki-Gun, Saitama, 350-0394, Japan (e-mail: otsuka@mail.dendai.ac.jp). Daiki Kakehi is graduate School of Science and Engineering, Tokyo Denki University. Hatoyama-Machi, Hiki-Gun, Saitama, 350-0394, Japan. Przemysław Ignaciuk is with Institute of Information Technology, Lodz University of Technology, al. Politechniki 8, 93-590Łódź, Poland (e-mail: przemyslaw.ignaciuk@p.lodz.pl).

Hence, the motivation for this study and the paper main contribution is to establish existence conditions and to design interval positive observers that would fulfill both objectives (i) and (ii). That is, we want to construct observers that use the output information in an efficient way and – at the same time – are robust against parametric uncertainties. Specifically, under the condition that all the subsystem matrices are subject to interval-type uncertainty and have identical output matrices, we give the existence conditions and design an $(n-p)$ dimensional interval reduced-order switched observers for continuous- and discrete-time linear positive switched systems under arbitrary switching, where $n$ is the dimension of state vector and $p$ is the dimension of measurement output. Furthermore, we study interval reduced-order positive observers for uncertain positive linear systems without switching as an important special case of practical systems.

The paper is organized in the following way: Section 2 introduces the principal assumptions and the design procedure of interval reduced-order positive observers for continuous-time uncertain switched positive linear systems. The discrete-time case is treated in Section 3. The analytical findings from Sections 2 and 3 are illustrated in a numerical study of two example systems in Section 4. Finally, Section 5 gives concluding remarks and discusses future prospects.

Throughout this study the following symbols and notation will be used:

· **N** – the set of natural numbers,

· **R** – the set of real numbers,

· **C** – the set of complex numbers,

· $I_n$ – an $n \times n$ identity matrix,

· $M > 0$ ($M \geq 0$) – a matrix whose $(i,j)$ elements $m_{ij}$ are all positive (nonnegative) for all $i,j$. Accordingly, $M$ is said to be positive (nonnegative);

· $M < 0$ ($M \leq 0$) – a matrix whose $(i,j)$ elements $m_{ij} < 0$ ($m_{ij} \leq 0$) for all $i,j$. Accordingly, $M$ is said to be negative (nonpositive);

· $\mathbf{R}_+^n$ – the nonnegative orthant of the $n$-dimensional real vector space $\mathbf{R}^n$,

· $\mathbf{R}_+^{n \times m}$ – a set of $n \times m$ nonnegative matrices,

· $M^{\mathrm{T}}$ – the transpose of matrix $M$,

· A matrix $M \in \mathbf{R}^{n \times n}$ is said to be Metzler if its off-diagonal elements are all nonnegative, i.e., $m_{ij} \geq 0$ for all $i \neq j$;

· A matrix $M \in \mathbf{R}^{n \times n}$ is said to be Hurwitz if all its eigenvalues are in the open left-region of **C**,

· A matrix $M \in \mathbf{R}^{n \times n}$ is said to be Schur if all its eigenvalues are in the open unit-disc of **C**,

· $\|x\|$ denotes the Euclidean norm of vector $x$,

· $\|M\|$ denotes the matrix norm of matrix $M \in \mathbf{R}^{n \times n}$.

## 2. CONTINUOUS-TIME INTERVAL REDUCED-ORDER POSITIVE OBSERVERS

Consider continuous-time switched linear systems of the form:

$$\Sigma_\sigma^c : \begin{cases} \dot{x}(t) = A_{\sigma(t)}x(t), & x(0) = x_0, \\ y(t) = Cx(t), \end{cases} \quad (1)$$

where $A_i \in \mathbf{R}^{n \times n}, x(t) \in \mathbf{R}^n$ is the state vector, $y(t) \in \mathbf{R}^p$ is the output and $\sigma(t) : [0,\infty) \to \{1,\ldots,N\}$ is the switching signal. In this study, it is assumed that the observation variables are the same for all the subsystems, i.e., the output matrix $C \in \mathbf{R}^{p \times n}$ has the form $C = [I_p \ 0] \geq 0$.

It is assumed that the initial state $x_0$ and the system matrices $A_i$ ($i = 1,\ldots,N$) are determined with limited precision with parameters in the range given by a set of interval-type inequalities:

**Assumption 1:**

(i) $0 \leq \underline{x}_0 \leq x_0 \leq \overline{x}_0$,

(ii) $\underline{A}_i \leq A_i \leq \overline{A}_i$ ($i = 1,\ldots,N$),

(iii) $\underline{A}_i$ ($i = 1,\ldots,N$) are Metzler,

where $\underline{x}_0, \overline{x}_0, \underline{A}_i$ and $\overline{A}_i$ are known vectors and matrices, respectively.

It follows from Assumption 1 that the uncertain switched system $\Sigma_\sigma^c$ is always positive. In order to design an $(n-p)$-dimensional switched positive observer for the system with $p$-dimensional measurement rank$C = p$, the initial vector and system matrices are decomposed as follows:

**Definition 1:**

(i) $\underline{x}_0 = \begin{bmatrix} \underline{x}_0^1 \\ \underline{x}_0^2 \end{bmatrix} \leq x_0 \leq \overline{x}_0 = \begin{bmatrix} \overline{x}_0^1 \\ \overline{x}_0^2 \end{bmatrix},$

(ii) $\underline{A}_i = \begin{bmatrix} \underline{A}_i^{11} & \underline{A}_i^{12} \\ \underline{A}_i^{21} & \underline{A}_i^{22} \end{bmatrix} \leq A_i = \begin{bmatrix} A_i^{11} & A_i^{12} \\ A_i^{21} & A_i^{22} \end{bmatrix}$

$\leq \overline{A}_i = \begin{bmatrix} \overline{A}_i^{11} & \overline{A}_i^{12} \\ \overline{A}_i^{21} & \overline{A}_i^{22} \end{bmatrix},$

where $\underline{x}_0^1, \overline{x}_0^1 \in \mathbf{R}^p$ and $\underline{A}_i^{11}, A_i^{11}, \overline{A}_i^{11} \in \mathbf{R}^{p \times p}$ ($i = 1,\ldots,N$).

Let $L \in \mathbf{R}_+^{(n-p) \times p}$ be a nonnegative matrix. For the purpose of observer design, the following auxiliary matrices are introduced:

**Definition 2:**

(i) $\underline{\hat{A}}_i := \underline{A}_i^{22} - L\overline{A}_i^{12}, \hat{A}_i := A_i^{22} - LA_i^{12}, \overline{\hat{A}}_i := \overline{A}_i^{22} - L\underline{A}_i^{12},$

(ii) $\underline{G}_i := \underline{\hat{A}}_i L + \underline{A}_i^{21} - L\overline{A}_i^{11}, \ G_i := \hat{A}_i L + A_i^{21} - LA_i^{11},$

$\overline{G}_i := \overline{\hat{A}}_i L + \overline{A}_i^{21} - L\underline{A}_i^{11},$

(iii) $F := \begin{bmatrix} -L & I_{n-p} \end{bmatrix}$, $\hat{C} := \begin{bmatrix} 0 \\ I_{n-p} \end{bmatrix} (\geq 0)$,

$$\hat{D} := \begin{bmatrix} I_p \\ L \end{bmatrix} (\geq 0).$$

Then, it follows from Assumption 1 (iii) that submatrices $\underline{A}_i^{11}$ and $\underline{A}_i^{22}$ ($i = 1, \ldots, N$) are Metzler and $\underline{A}_i^{12}$ and $\underline{A}_i^{21}$ ($i = 1, \ldots, N$) are nonnegative. Further, since $L$ is nonnegative, the following inequalities are easily obtained:

$$\underline{\hat{A}}_i \leq \hat{A}_i \leq \overline{\hat{A}}_i, \ \underline{G}_i \leq G_i \leq \overline{G}_i \ (i = 1, \ldots, N). \quad (2)$$

In order to demonstrate the main result stated below in Theorem 1, let us first cite two lemmas.

**Lemma 1:** [3]
A matrix $M$ is Metzler if and only if $e^{Mt} \geq 0$ for all $t \geq 0$.

**Lemma 2:** [12]
If two matrices $M$ and $N$ are Metzler satisfying $M \leq N$, then $(0 \leq) e^{Mt} \leq e^{Nt}$ for all $t \geq 0$.

In this work, it is assumed that the states of switched systems and their estimates provided by the reduced-order switched observers do not jump at the switching instants, i.e., the state continuity at the switching instants is preserved. Note that if the observer gain $L$ depends on $i$, where $i$ is the subsystem number, then $\hat{D}$ also depends on $i$, which means that the continuity of the state estimates does not hold in general. Therefore, it is assumed here that the observer gain $L$ does not depend on $i$.

The main result of this section is given in the following theorem.

**Theorem 1:** Suppose that the state trajectory $x(t)$ of the uncertain switched positive linear system $\Sigma_\sigma^c$ is bounded and Assumption 1 holds. If there exists an $L \in \mathbf{R}_+^{(n-p) \times p}$ such that the following conditions:

(i) $\underline{\hat{A}}_i$ ($i = 1, \ldots, N$) are Metzler,

(ii) $\underline{G}_i \geq 0$ ($i = 1, \ldots, N$),

(iii) There exists a $\lambda > 0$
such that $(\overline{\hat{A}}_i)^T \lambda < 0$ ($i = 1, \ldots, N$),

(iv) $\omega_0^\ell \leq \underline{x}_0^2 - L\overline{x}_0^1$ and $\overline{x}_0^2 - L\underline{x}_0^1 \leq \omega_0^u$

are satisfied, then there exists an interval reduced-order switched positive observer $\left(\hat{\Sigma}_\sigma^{cr\ell}, \hat{\Sigma}_\sigma^{cru}\right)$ with the initial states $\omega_0^\ell$ and $\omega_0^u$ satisfying $\omega_0^\ell \leq \omega_0^u$ for $\Sigma_\sigma^c$ such that $\xi(t) (:= \hat{x}^u(t) - \hat{x}^\ell(t))$ is bounded and

$$0 \leq \hat{x}^\ell(t) \leq x(t) \leq \hat{x}^u(t) \ (t \geq 0)$$

under arbitrary switching, where

$$\hat{\Sigma}_\sigma^{cr\ell} : \begin{cases} \dot{\omega}^\ell(t) = \underline{\hat{A}}_{\sigma(t)} \omega^\ell(t) + \underline{G}_{\sigma(t)} y(t), \ \omega^\ell(0) = \omega_0^\ell (\geq 0), \\ \hat{x}^\ell(t) := \hat{C} \omega^\ell(t) + \hat{D} y(t) \end{cases}$$

and

$$\hat{\Sigma}_\sigma^{cru} : \begin{cases} \dot{\omega}^u(t) = \overline{\hat{A}}_{\sigma(t)} \omega^u(t) + \overline{G}_{\sigma(t)} y(t), \ \omega^u(0) = \omega_0^u (\geq 0), \\ \hat{x}^u(t) := \hat{C} \omega^u(t) + \hat{D} y(t). \end{cases}$$

Here, $\omega^\ell(t)$ and $\omega^u(t)$ are lower and upper observer states, and $\hat{x}^\ell(t)$ and $\hat{x}^u(t)$ are lower and upper state estimates, respectively. Matrix $L \in \mathbf{R}_+^{(n-p) \times p}$ is the observer gain for interval reduced order switched positive observer $\left(\hat{\Sigma}_\sigma^{cr\ell}, \hat{\Sigma}_\sigma^{cru}\right)$.

**Proof:** Suppose there exists an $L \in \mathbf{R}_+^{(n-p) \times p}$ such that conditions (i)-(iv) are satisfied. Then, one obtains from condition (iv): $\omega_0^\ell \leq \omega_0^u$. Now, let us consider the lower and upper $(n - p)$-dimensional switched systems $\left(\hat{\Sigma}_\sigma^{cr\ell}, \hat{\Sigma}_\sigma^{cru}\right)$ together with relations (2). In order to show that $\hat{x}^\ell(t) \leq x(t) \leq \hat{x}^u(t)$ ($t \geq 0$), let us introduce two uncertain switched systems:

$$\underline{\hat{\Sigma}}_\sigma^{cr} : \begin{cases} \dot{\underline{\omega}}(t) = \hat{A}_{\sigma(t)} \underline{\omega}(t) + G_{\sigma(t)} y(t), \ \underline{\omega}(0) = \omega_0^\ell (\geq 0) \\ \underline{\hat{x}}(t) := \hat{C} \underline{\omega}(t) + \hat{D} y(t). \end{cases}$$

$$\overline{\hat{\Sigma}}_\sigma^{cr} : \begin{cases} \dot{\overline{\omega}}(t) = \hat{A}_{\sigma(t)} \overline{\omega}(t) + G_{\sigma(t)} y(t), \ \overline{\omega}(0) = \omega_0^u (\geq 0) \\ \overline{\hat{x}}(t) := \hat{C} \overline{\omega}(t) + \hat{D} y(t). \end{cases}$$

Suppose that $\{0, t_1, \cdots, t_k, \cdots\}$ be arbitrary fixed switching time sequence of $\sigma$ in $[0, \infty)$ and $\{i_0 := \sigma(0+), i_1 := \sigma(t_1+), \cdots, i_k := \sigma(t_k+), \cdots\}$ be the switching index sequence of $\sigma$ in $[0, \infty)$. Then, we have the following solutions for the above four switched systems:

$$\omega^\ell(t) = e^{\underline{\hat{A}}_{i_k}(t-t_k)} e^{\underline{\hat{A}}_{i_{k-1}}(t_k-t_{k-1})} \cdots e^{\underline{\hat{A}}_{i_0}(t_1-t_0)} \omega_0^\ell$$
$$+ e^{\underline{\hat{A}}_{i_k}(t-t_k)} \cdots e^{\underline{\hat{A}}_{i_1}(t_2-t_1)} \int_0^{t_1} e^{\underline{\hat{A}}_{i_0}(t_1-\tau)} \underline{G}_{i_0} y(\tau) d\tau$$
$$+ \cdots + e^{\underline{\hat{A}}_{i_k}(t-t_k)} \int_{t_{k-1}}^{t_k} e^{\underline{\hat{A}}_{i_{k-1}}(t_k-\tau)} \underline{G}_{i_{k-1}} y(\tau) d\tau$$
$$+ \int_{t_k}^{t} e^{\underline{\hat{A}}_{i_k}(t-\tau)} \underline{G}_{i_k} y(\tau) d\tau.$$

$$\underline{\omega}(t) = e^{\hat{A}_{i_k}(t-t_k)} e^{\hat{A}_{i_{k-1}}(t_k-t_{k-1})} \cdots e^{\hat{A}_{i_0}(t_1-t_0)} \omega_0^\ell$$
$$+ e^{\hat{A}_{i_k}(t-t_k)} \cdots e^{\hat{A}_{i_1}(t_2-t_1)} \int_0^{t_1} e^{\hat{A}_{i_0}(t_1-\tau)} G_{i_0} y(\tau) d\tau$$
$$+ \cdots + e^{\hat{A}_{i_k}(t-t_k)} \int_{t_{k-1}}^{t_k} e^{\hat{A}_{i_{k-1}}(t_k-\tau)} G_{i_{k-1}} y(\tau) d\tau$$
$$+ \int_{t_k}^{t} e^{\hat{A}_{i_k}(t-\tau)} G_{i_k} y(\tau) d\tau.$$

$$\overline{\omega}(t) = e^{\hat{A}_{i_k}(t-t_k)} e^{\hat{A}_{i_{k-1}}(t_k-t_{k-1})} \cdots e^{\hat{A}_{i_0}(t_1-t_0)} \omega_0^u$$
$$+ e^{\hat{A}_{i_k}(t-t_k)} \cdots e^{\hat{A}_{i_1}(t_2-t_1)} \int_0^{t_1} e^{\hat{A}_{i_0}(t_1-\tau)} G_{i_0} y(\tau) d\tau$$
$$+ \cdots + e^{\hat{A}_{i_k}(t-t_k)} \int_{t_{k-1}}^{t_k} e^{\hat{A}_{i_{k-1}}(t_k-\tau)} G_{i_{k-1}} y(\tau) d\tau$$
$$+ \int_{t_k}^{t} e^{\hat{A}_{i_k}(t-\tau)} G_{i_k} y(\tau) d\tau.$$

$$\omega^u(t) = e^{\hat{\bar{A}}_{i_k}(t-t_k)} e^{\hat{\bar{A}}_{i_{k-1}}(t_k-t_{k-1})} \cdots e^{\hat{\bar{A}}_{i_0}(t_1-t_0)} \omega_0^u$$
$$+ e^{\hat{\bar{A}}_{i_k}(t-t_k)} \cdots e^{\hat{\bar{A}}_{i_1}(t_2-t_1)} \int_0^{t_1} e^{\hat{\bar{A}}_{i_0}(t_1-\tau)} \overline{G}_{i_0} y(\tau) d\tau$$
$$+ \cdots + e^{\hat{\bar{A}}_{i_k}(t-t_k)} \int_{t_{k-1}}^{t_k} e^{\hat{\bar{A}}_{i_{k-1}}(t_k-\tau)} \overline{G}_{i_{k-1}} y(\tau) d\tau$$
$$+ \int_{t_k}^{t} e^{\hat{\bar{A}}_{i_k}(t-\tau)} \overline{G}_{i_k} y(\tau) d\tau.$$

Since $\omega_0^\ell \leq \omega_0^u$, using conditions (i), (ii), and (2) together with Lemma 1 and 2, the following set of inequalities can be formulated:

$$(0 \leq) \; \omega^\ell(t) \leq \underline{\omega}(t) \leq \overline{\omega}(t) \leq \omega^u(t) \text{ and} \tag{3}$$

$$(0 \leq) \; \hat{x}^\ell(t) \leq \underline{\hat{x}}(t) \leq \hat{\bar{x}}(t) \leq \hat{x}^u(t). \tag{4}$$

Define the error as $\underline{\epsilon}(t) := Fx(t) - \underline{\omega}(t)$. Then, we have

$$\begin{aligned}
\underline{\dot{\epsilon}}(t) &= F\dot{x}(t) - \underline{\dot{\omega}}(t) \\
&= (FA_\sigma - G_\sigma C) x(t) - \hat{A}_\sigma \underline{\omega}(t) \\
&= \hat{A}_\sigma \left( Fx(t) - \underline{\omega}(t) \right) \\
&= \hat{A}_\sigma \underline{\epsilon}(t),
\end{aligned} \tag{5}$$

and the initial error

$$\underline{\epsilon}(0) = Fx(0) - \underline{\omega}(0) = x_0^2 - Lx_0^1 - \omega_0^\ell.$$

It follows from condition (iv) that $\underline{\epsilon}(0) \geq 0$. Since $\hat{A}_i$ in (5) is Metzler, by virtue of condition (i) and $\underline{\epsilon}(0) \geq 0$, we have $\underline{\epsilon}(t) \geq 0$, and

$$Fx(t) \geq \underline{\omega}(t). \tag{6}$$

Similarly, with the following error $\overline{\epsilon}(t) := \overline{\omega}(t) - Fx(t)$ one obtains from (i) and (iv) that

$$Fx(t) \leq \overline{\omega}(t). \tag{7}$$

Hence, it follows from (6) and (7) that

$$\underline{\omega}(t) \leq Fx(t) \leq \overline{\omega}(t) \; (t \geq 0). \tag{8}$$

We are now ready to prove the following claim.

*Claim* 1: $\underline{\hat{x}}(t) \leq x(t) \leq \hat{\bar{x}}(t) \; (t \geq 0)$.

*Proof of Claim* 1:
$$\begin{aligned}
x(t) - \underline{\hat{x}}(t) &= x(t) - \hat{C}\underline{\omega}(t) - \hat{D}y(t) \\
&= (I_n - \hat{D}C) x(t) - \hat{C}\underline{\omega}(t) \\
&= \left( \begin{bmatrix} I_p & 0 \\ 0 & I_{n-p} \end{bmatrix} - \begin{bmatrix} I_p & 0 \\ L & 0 \end{bmatrix} \right) x(t) - \hat{C}\underline{\omega}(t) \\
&= \begin{bmatrix} 0 & 0 \\ -L & I_{n-p} \end{bmatrix} x(t) - \begin{bmatrix} 0 \\ I_{n-p} \end{bmatrix} \underline{\omega}(t) \\
&= \begin{bmatrix} 0 \\ Fx(t) - \underline{\omega}(t) \end{bmatrix} \geq 0,
\end{aligned}$$

which implies that $\underline{\hat{x}}(t) \leq x(t)$. Similarly,

$$\begin{aligned}
\hat{\bar{x}}(t) - x(t) &= \hat{C}\overline{\omega}(t) + \hat{D}y(t) - x(t) \\
&= (\hat{D}C - I_n) x(t) + \hat{C}\overline{\omega}(t) \\
&= \left( \begin{bmatrix} I_p & 0 \\ L & 0 \end{bmatrix} - \begin{bmatrix} I_p & 0 \\ 0 & I_{n-p} \end{bmatrix} \right) x(t) + \hat{C}\overline{\omega}(t) \\
&= \begin{bmatrix} 0 & 0 \\ L & -I_{n-p} \end{bmatrix} x(t) + \begin{bmatrix} 0 \\ I_{n-p} \end{bmatrix} \overline{\omega}(t) \\
&= \begin{bmatrix} 0 \\ \overline{\omega}(t) - Fx(t) \end{bmatrix} \geq 0,
\end{aligned}$$

which implies that $x(t) \leq \hat{\bar{x}}(t)$. Thus, Claim 1 was shown true.

It follows from (4) and Claim 1 that

$$(0 \leq) \; \hat{x}^\ell(t) \leq x(t) \leq \hat{x}^u(t) \; (t \geq 0). \tag{9}$$

Next, we prove the following claim.

*Claim* 2: $\xi(t) := \hat{x}^u(t) - \hat{x}^\ell(t)$ is bounded.

*Proof of Claim* 2: Since

$$\begin{aligned}
\xi(t) &= \hat{x}^u(t) - \hat{x}^\ell(t) \\
&= \hat{C}\omega^u(t) + \hat{D}Cx(t) - \hat{C}\omega^\ell(t) - \hat{D}Cx(t) \\
&= \hat{C}\left(\omega^u(t) - \omega^\ell(t)\right),
\end{aligned}$$

we have

$$\|\xi(t)\| \leq \|\hat{C}\| \cdot \|\omega^u(t) - \omega^\ell(t)\|. \tag{10}$$

Since the trajectory of $x(t)$ is bounded and $\omega^u(t)$ is asymptotically stable under arbitrary switching, from (iii), the trajectory of $\omega^u(t)$ is bounded which implies $\omega^\ell(t)$ is also bounded. Hence, it follows from (10) that $\xi(t)$ is bounded. Consequently, Claim 2 holds true. This conclusion ends the proof of Theorem 1. □

The design procedure of the proposed interval reduced-order switched positive observer can be summarized in the following algorithm:

**STEP 1.** Obtain the state dimension $n$ and rank $p$ of the output matrix $C = [I_p \; 0]$.

**STEP 2.** Partition matrices $\overline{A}_i$, $\underline{A}_i$, and vectors $\overline{x}_0$ and $\underline{x}_0$ according to Definition 1.

**STEP 3.** Provide the observer initial states $\omega_0^\ell$ and $\omega_0^u$ satisfying $0 \leq \omega_0^\ell \leq \omega_0^u$.

**STEP 4.** Calculate the observer gain $L \in \mathbf{R}_+^{(n-p) \times p}$ satisfying conditions (i)-(iv) of Theorem 1.

**STEP 5.** Calculate $\hat{\overline{A}}_i$, $\hat{\underline{A}}_i$, $\overline{G}_i$, $\underline{G}_i$, $F$, $\hat{C}$ and $\hat{D}$ using the observer gain $L$ obtained in STEP 4 according to Definition 2.

**STEP 6.** Calculate the parameters of interval reduced-order switched positive observer $(\hat{\Sigma}_\sigma^{cr\ell}, \hat{\Sigma}_\sigma^{cru})$ using the matrices obtained in STEP 5 and the initial states $\omega_0^\ell$ and $\omega_0^u$ obtained in STEP 3 according to the formulas from Theorem 1.

Note that the reduced-order interval observer obtained from STEP1-STEP6 is robust against uncertainty on the initial state $x_0$ and interval matrices $A_i$ $(i = 1, \ldots, N)$.

Now, we will consider a special case when the switching signal $\sigma(t)$ for the uncertain switched linear system $\Sigma_\sigma^c$ is constant. In other words, we will consider the following uncertain continuous-time *non-switched* linear system:

$$\Sigma^c : \begin{cases} \dot{x}(t) = Ax(t), & x(0) = x_0, \\ y(t) = Cx(t), \end{cases} \quad (11)$$

satisfying Assumption 1, where $A := A_1 = A_2 = \cdots = A_N \in \mathbf{R}^{n \times n}$, $x(t) \in \mathbf{R}^n$ is the state vector, and $y(t) \in \mathbf{R}^p$ is the output.

For the systems described by (11), with interval-type uncertainties on the initial state, system matrix $A$ and output matrix $C$, Rami et al. [12] specifies existence conditions of a full-order interval positive observer and discusses their feasibility. The following corollary provides existence conditions of a *reduced*-order interval positive observer for this class of systems with the output matrix $C = [I_p\ 0]\ (\geq 0)$.

**Corollary 1:** Suppose the state trajectory $x(t)$ of the uncertain positive linear system $\Sigma^c$ (11) is bounded and Assumption 1 holds. If there exists an $L \in \mathbf{R}_+^{(n-p) \times p}$ such that the following conditions :

(i) $\underline{\hat{A}}$ is Metzler,

(ii) $\underline{G} \geq 0$,

(iii) $\hat{\overline{A}}$ is Hurwitz,

(iv) $\omega_0^\ell \leq \underline{x}_0^2 - L\overline{x}_0^1$ and $\overline{x}_0^2 - L\underline{x}_0^1 \leq \omega_0^u$

are satisfied, then there exists an interval reduced-order positive observer $(\hat{\Sigma}^{cr\ell}, \hat{\Sigma}^{cru})$ with the initial states $\omega_0^\ell$ and $\omega_0^u$ satisfying $\omega_0^\ell \leq \omega_0^u$ for $\Sigma^c$ such that $\xi(t)(:= \hat{x}^u(t) - \hat{x}^\ell(t))$ is bounded and

$$0 \leq \hat{x}^\ell(t) \leq x(t) \leq \hat{x}^u(t),$$

where

$$\hat{\Sigma}^{cr\ell} : \begin{cases} \dot{\omega}^\ell(t) = \underline{\hat{A}}\omega^\ell(t) + \underline{G}y(t), & \omega^\ell(0) = \omega_0^\ell (\geq 0), \\ \hat{x}^\ell(t) := \hat{C}\omega^\ell(t) + \hat{D}y(t) \end{cases}$$

and

$$\hat{\Sigma}^{cru} : \begin{cases} \dot{\omega}^u(t) = \hat{\overline{A}}\omega^u(t) + \overline{G}y(t), & \omega^u(0) = \omega_0^u (\geq 0), \\ \hat{x}^u(t) := \hat{C}\omega^u(t) + \hat{D}y(t). \end{cases}$$

$\omega^\ell(t)$ and $\omega^u(t)$ are lower and upper observer states, and $\hat{x}^\ell(t)$ and $\hat{x}^u(t)$ are the lower and upper state estimates, respectively. $L \in \mathbf{R}_+^{(n-p) \times p}$ is the observer gain for the interval reduced order positive observer $(\hat{\Sigma}^{cr\ell}, \hat{\Sigma}^{cru})$.

## 3. DISCRETE-TIME INTERVAL REDUCED-ORDER POSITIVE OBSERVERS

Consider the following discrete-time switched linear system

$$\Sigma_\sigma^d : \begin{cases} x(k+1) = A_{\sigma(k)}x(k), & x(0) = x_0, \\ y(k) = Cx(k), \end{cases} \quad (12)$$

where $A_i \in \mathbf{R}^{n \times n}$, $x(k) \in \mathbf{R}^n$ is the state vector, $y(k) \in \mathbf{R}^p$ is the output and $\sigma(k) : [0, 1, 2, \ldots) \to \{1, \ldots, N\}$ is the switching signal. The output matrix $C \in \mathbf{R}^{p \times n}$ is of the form $C = [I_p\ 0] \geq 0$. It is assumed that the initial state $x_0$ and system matrices $A_i$ $(i = 1, \ldots, N)$ are subject to the following interval-type uncertainty:

**Assumption 2:**

(i) $0 \leq \underline{x}_0 \leq x_0 \leq \overline{x}_0$,

(ii) $\underline{A}_i \leq A_i \leq \overline{A}_i\ (i = 1, \ldots, N)$,

(iii) $\underline{A}_i \geq 0\ (i = 1, \ldots, N)$,

where $\underline{x}_0, \overline{x}_0, \underline{A}_i$ and $\overline{A}_i$ are known vectors and matrices.

Using the partitioning scheme introduced in Definitions 1 and 2 in the previous section, the following theorem can be established. Since, in its essence, the reasoning is similar to the continuous-time case, the proof will be omitted.

**Theorem 2:** Suppose that the state trajectory $x(k)$ of the uncertain switched positive linear system $\Sigma_\sigma^d$ is bounded and Assumption 2 holds. If there exists an $L \in \mathbf{R}_+^{(n-p) \times p}$ such that the following conditions :

(i) $\underline{\hat{A}}_i \geq 0\ (i = 1, \ldots, N)$,

(ii) $\underline{G}_i \geq 0\ (i = 1, \ldots, N)$,

(iii) There exists a $\lambda > 0$

such that $(\hat{\overline{A}}_i - I_{n-p})^\mathrm{T}\lambda < 0\ (i = 1, \ldots, N)$,

(iv) $\omega_0^\ell \leq \underline{x}_0^2 - L\overline{x}_0^1$ and $\overline{x}_0^2 - L\underline{x}_0^1 \leq \omega_0^u$

are satisfied, then there exists an interval reduced-order switched positive observer $(\hat{\Sigma}_\sigma^{dr\ell}, \hat{\Sigma}_\sigma^{dru})$ with the initial states $\omega_0^\ell$ and $\omega_0^u$ satisfying $\omega_0^\ell \leq \omega_0^u$ for $\Sigma_\sigma^d$ such that $\xi(k)(:= \hat{x}^u(k) - \hat{x}^\ell(k))$ is bounded and

$$0 \leq \hat{x}^\ell(k) \leq x(k) \leq \hat{x}^u(k)\ (k = 0, 1, \ldots)$$

under arbitrary switching, where

$$\hat{\Sigma}_\sigma^{dr\ell} : \begin{cases} \omega^\ell(k+1) = \underline{\hat{A}}_{\sigma(k)}\omega^\ell(k) + \underline{G}_{\sigma(k)}y(k), \ \omega^\ell(0) = \omega_0^\ell, \\ \hat{x}^\ell(k) = \hat{C}\omega^\ell(k) + \hat{D}y(k) \end{cases}$$

and

$$\hat{\Sigma}_\sigma^{dru} : \begin{cases} \omega^u(k+1) = \overline{\hat{A}}_{\sigma(k)}\omega^u(k) + \overline{G}_{\sigma(k)}y(k), \ \omega^u(0) = \omega_0^u, \\ \hat{x}^u(k) = \hat{C}\omega^u(k) + \hat{D}y(k). \end{cases}$$

Functions $\omega^\ell(k)$ and $\omega^u(k)$ represent the lower and upper observer states, and $\hat{x}^\ell(k)$ and $\hat{x}^u(k)$ are the lower and upper state estimates, respectively. $L \in \mathbf{R}_+^{(n-p) \times p}$ is the observer gain for the interval reduced order switched positive observer $\left(\hat{\Sigma}_\sigma^{dr\ell}, \hat{\Sigma}_\sigma^{dru}\right)$.

Let us now consider a special case when the switching signal $\sigma(k)$ for the uncertain switched linear system $\Sigma_\sigma^d$ is constant, i.e., we will analyze the following discrete-time non-switched linear system:

$$\Sigma^d : \begin{cases} x(k+1) = Ax(k), \quad x(0) = x_0, \\ y(k) = Cx(k), \end{cases} \quad (13)$$

satisfying Assumption 2, where $A := A_1 = A_2 = \cdots = A_N \in \mathbf{R}^{n \times n}$, $x(k) \in \mathbf{R}^n$ is the state vector, and $y(k) \in \mathbf{R}^p$ is the output of the form $C = [I_p \ 0] \ (\geq 0)$.

The corollary formulated below gives existence conditions of a *reduced*-order interval positive observer for discrete-time uncertain positive linear systems (13).

**Corollary 2:** Suppose that the state trajectory $x(k)$ of the uncertain positive system $\Sigma^d$ is bounded and Assumption 2 holds. If there exists an $L \in \mathbf{R}_+^{(n-p) \times p}$ such that the following conditions :

(i) $\underline{\hat{A}} \geq 0$,

(ii) $\underline{G} \geq 0$,

(iii) $\overline{\hat{A}}$ is Schur,

(iv) $\omega_0^\ell \leq \underline{x}_0^2 - L\overline{x}_0^1$ and $\overline{x}_0^2 - L\underline{x}_0^1 \leq \omega_0^u$

are satisfied, then there exists an interval reduced-order positive observer $\left(\hat{\Sigma}^{dr\ell}, \hat{\Sigma}^{dru}\right)$ with the initial states $\omega_0^\ell$ and $\omega_0^u$ satisfying $\omega_0^\ell \leq \omega_0^u$ for $\Sigma^d$ such that $\xi(k)(:= \hat{x}^u(k) - \hat{x}^\ell(k))$ is bounded and

$$0 \leq \hat{x}^\ell(k) \leq x(k) \leq \hat{x}^u(k) \ (k = 0, 1, \ldots),$$

where

$$\hat{\Sigma}^{dr\ell} : \begin{cases} \omega^\ell(k+1) = \underline{\hat{A}}\omega^\ell(k) + \underline{G}y(k), \quad \omega^\ell(0) = \omega_0^\ell, \\ \hat{x}^\ell(k) = \hat{C}\omega^\ell(k) + \hat{D}y(k) \end{cases}$$

and

$$\hat{\Sigma}^{dru} : \begin{cases} \omega^u(k+1) = \overline{\hat{A}}\omega^u(k) + \overline{G}y(k), \quad \omega^u(0) = \omega_0^u, \\ \hat{x}^u(k) = \hat{C}\omega^u(k) + \hat{D}y(k). \end{cases}$$

Functions $\omega^\ell(k)$ and $\omega^u(k)$ reflect the lower and upper observer states, and $\hat{x}^\ell(k)$ and $\hat{x}^u(k)$ are the lower and upper state estimates, respectively. $L \in \mathbf{R}_+^{(n-p) \times p}$ is the observer gain for the interval reduced order positive observer $\left(\hat{\Sigma}^{dr\ell}, \hat{\Sigma}^{dru}\right)$.

## 4. EXAMPLES

In this section, two numerical examples illustrating properties of the designed continuous- and discrete-time interval reduced-order switched positive observers will be presented.

### 4.1. Continuous-Time Case

Consider a five-dimensional continuous-time switched positive linear system comprising three subsystems with interval uncertainty on its parameters and a common output matrix $C$ of rank$C = p = 2$. The uncertainty bounds are specified by the following matrices and vectors:

$$\underline{A}_1 = \begin{bmatrix} -23 & 4 & 1 & 4 & 1 \\ 6 & -28 & 8 & 6 & 8 \\ \hline 4 & 4 & -25 & 4 & 6 \\ 7 & 5 & 6 & -26 & 3 \\ 5 & 4 & 2 & 6 & -29 \end{bmatrix},$$

$$\overline{A}_1 = \begin{bmatrix} -22 & 5 & 2 & 5 & 3 \\ 8 & -26 & 9 & 7 & 9 \\ \hline 7 & 6 & -24 & 6 & 7 \\ 8 & 6 & 8 & -24 & 4 \\ 8 & 5 & 5 & 7 & -27 \end{bmatrix},$$

$$\underline{A}_2 = \begin{bmatrix} -28 & 6 & 3 & 1 & 4 \\ 1 & -24 & 6 & 3 & 1 \\ \hline 6 & 5 & -29 & 8 & 3 \\ 4 & 8 & 3 & -23 & 2 \\ 3 & 4 & 2 & 2 & -22 \end{bmatrix},$$

$$\overline{A}_2 = \begin{bmatrix} -26 & 9 & 4 & 2 & 5 \\ 3 & -22 & 7 & 4 & 3 \\ \hline 8 & 7 & -27 & 10 & 6 \\ 6 & 10 & 7 & -19 & 4 \\ 5 & 6 & 4 & 3 & -18 \end{bmatrix},$$

$$\underline{A}_3 = \begin{bmatrix} -28 & 10 & 1 & 4 & 4 \\ 1 & -27 & 8 & 6 & 3 \\ \hline 2 & 4 & -29 & 8 & 3 \\ 4 & 2 & 5 & -26 & 9 \\ 1 & 3 & 8 & 5 & -28 \end{bmatrix},$$

$$\overline{A}_3 = \begin{bmatrix} -26 & 13 & 3 & 6 & 6 \\ 3 & -25 & 9 & 7 & 4 \\ \hline 4 & 7 & -26 & 10 & 4 \\ 6 & 4 & 7 & -24 & 12 \\ 2 & 5 & 9 & 6 & -26 \end{bmatrix},$$

$$\underline{x}_0 = \begin{bmatrix} 1 \\ 3 \\ 6 \\ 2 \\ 3 \end{bmatrix}, \quad \overline{x}_0 = \begin{bmatrix} 6 \\ 5 \\ 9 \\ 8 \\ 5 \end{bmatrix}, \quad C = \begin{bmatrix} 1 & 0 & 0 & 0 & 0 \\ 0 & 1 & 0 & 0 & 0 \end{bmatrix}.$$

Since the first two components of the state vector – $x_1(t)$ and $x_2(t)$ – can be directly read as the system output $y(t)$, in the experiments, the third, fourth, and fifth component of $x(t)$ will be estimated. In the design of interval reduced-order observers, we will follow the steps of the procedure defined in Section 2.

**STEP 1:** The system dimension $n = 5$, the output dimension $p = 2$.

**STEP 2:** The matrices grouping the subsystem parameters and the initial state vector are partitioned as:

$$A_1 = \begin{bmatrix} -22.13 & 4.64 & 1.76 & 4.21 & 1.08 \\ 7.18 & -26.20 & 8.40 & 6.08 & 8.90 \\ \hline 6.31 & 4.66 & -24.04 & 5.18 & 6.18 \\ 7.75 & 5.49 & 7.22 & -25.48 & 3.72 \\ 5.39 & 4.23 & 3.05 & 6.58 & -28.76 \end{bmatrix},$$

$$A_2 = \begin{bmatrix} -27.04 & 6.72 & 3.97 & 1.61 & 4.36 \\ 2.48 & -22.74 & 6.00 & 3.12 & 1.18 \\ \hline 7.06 & 6.50 & -28.84 & 9.16 & 3.54 \\ 5.28 & 9.70 & 5.28 & -19.52 & 3.22 \\ 4.92 & 5.34 & 2.68 & 2.13 & -18.28 \end{bmatrix},$$

$$A_3 = \begin{bmatrix} -26.52 & 10.09 & 2.06 & 5.98 & 4.76 \\ 1.02 & -25.66 & 8.05 & 6.85 & 3.03 \\ \hline 2.90 & 6.04 & -28.31 & 9.86 & 3.77 \\ 5.10 & 2.76 & 5.86 & -24.26 & 10.05 \\ 1.36 & 3.86 & 8.14 & 5.39 & -27.60 \end{bmatrix},$$

$$x_0 = \begin{bmatrix} 4.45 \\ 3.42 \\ 6.33 \\ 7.64 \\ 4.72 \end{bmatrix}.$$

Note that since $\underline{A}_i$ ($i = 1,2,3$) are Metzler and $\underline{x}_0 \geq 0$, all the conditions stated in Assumption 1 are fulfilled.

**STEP 3:** The initial observer states $\omega_0^\ell$ and $\omega_0^u$ are chosen as:

$$\omega_0^\ell = \begin{bmatrix} 1 \\ 0 \\ 1 \end{bmatrix}, \quad \omega_0^u = \begin{bmatrix} 8 \\ 8 \\ 9 \end{bmatrix}.$$

**STEP 4:** The gain of the interval reduced-order observer, selected as

$$L = \begin{bmatrix} 0.1 & 0.4 \\ 0.15 & 0.2 \\ 0.1 & 0.05 \end{bmatrix} \in \mathbf{R}_+^{3 \times 2}$$

satisfies conditions (i)-(iv) of Theorem 1.

**STEP 5 and 6:**

It follows from Theorem 1 that there exists an interval reduced-order switched positive observer $\left(\hat{\Sigma}_\sigma^{cr\ell}, \hat{\Sigma}_\sigma^{cru}\right)$ with $\omega_0^\ell \leq \omega_0^u$ for $\Sigma_\sigma^c$ such that $\xi(t) (:= \hat{x}^u(t) - \hat{x}^\ell(t))$ is bounded and

$$0 \leq \hat{x}^\ell(t) \leq x(t) \leq \hat{x}^u(t) \ (t \geq 0)$$

under arbitrary switching, where

$$\hat{\Sigma}_\sigma^{cr\ell} : \begin{cases} \dot{\omega}^\ell(t) = \underline{\hat{A}}_{\sigma(t)} \omega^\ell(t) + \underline{G}_{\sigma(t)} y(t), \ \omega^\ell(0) = \omega_0^\ell (\geq 0), \\ \hat{x}^\ell(t) := \hat{C} \omega^\ell(t) + \hat{D} y(t), \end{cases}$$

and

$$\hat{\Sigma}_\sigma^{cru} : \begin{cases} \dot{\omega}^u(t) = \hat{\overline{A}}_{\sigma(t)} \omega^u(t) + \overline{G}_{\sigma(t)} y(t), \ \omega^u(0) = \omega_0^u (\geq 0), \\ \hat{x}^u(t) := \hat{C} \omega^u(t) + \hat{D} y(t). \end{cases}$$

Moreover, based on the findings from the proof of Theorem 1, the following two uncertain interval reduced-order observers can be formulated:

$$\underline{\hat{\Sigma}}_\sigma^{cr} : \begin{cases} \underline{\dot{\omega}}(t) = \hat{A}_{\sigma(t)} \underline{\omega}(t) + G_{\sigma(t)} y(t), \ \underline{\omega}(0) = \omega_0^\ell (\geq 0), \\ \underline{\hat{x}}(t) := \hat{C} \underline{\omega}(t) + \hat{D} y(t), \end{cases}$$

$$\hat{\overline{\Sigma}}_\sigma^{cr} : \begin{cases} \dot{\overline{\omega}}(t) = \hat{A}_{\sigma(t)} \overline{\omega}(t) + G_{\sigma(t)} y(t), \ \overline{\omega}(0) = \omega_0^u (\geq 0), \\ \hat{\overline{x}}(t) := \hat{C} \overline{\omega}(t) + \hat{D} y(t). \end{cases}$$

The results of Matlab simulations are illustrated in Figs. 1 and 2. Figure 1 depicts the state trajectory and its estimate performed by the designed observer and Fig. 2 shows the switching signal. We can see from Fig. 1 that the state $x(t)$ is within the observer estimates $\hat{x}^\ell(t)$ and $\hat{x}^u(t)$,

$$0 \leq \hat{x}^\ell(t) \leq x(t) \leq \hat{x}^u(t) \ (t \geq 0),$$

under a random switching (Fig. 2). The chosen form of the output matrix of $\mathrm{rank} C = p = 2$ gives the state variables $x_1(t)$ and $x_2(t)$ equal to the first and the second elements of the lower bound state estimate $\hat{x}^\ell(t)$ and the upper bound state estimate $\hat{x}^u(t)$, respectively. The remaining $(5 - p) = 3$ elements of the state are correctly estimated by the designed interval reduced-order observer with the estimation error asymptotically driven to zero.

### 4.2. Discrete-Time Case

Consider a four-dimensional discrete-time switched positive linear system consisting of three subsystems with interval-type uncertainty and a common output matrix $C$ of $\mathrm{rank} C = p = 2$. The uncertainty range is given by the following matrices and vectors:

$$\underline{A}_1 = \begin{bmatrix} 0.03 & 0.07 & 0.01 & 0.14 \\ 0.12 & 0.13 & 0.02 & 0.08 \\ \hline 0.07 & 0.03 & 0.01 & 0.04 \\ 0.08 & 0.02 & 0.21 & 0.11 \end{bmatrix},$$

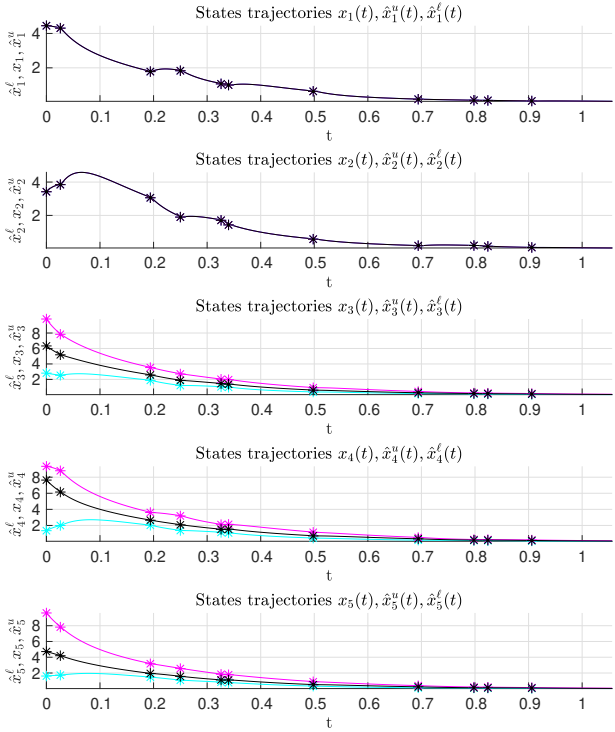

Fig. 1. System state $x(t)$ and estimates $\hat{x}^\ell(t)$, $\hat{x}^u(t)$ provided by the interval reduced-order switched positive observer – continuous-time case

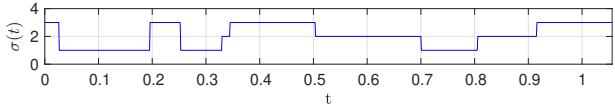

Fig. 2. Switching signal – continuous-time case

$$\overline{A}_1 = \left[\begin{array}{cc|cc} 0.20 & 0.15 & 0.30 & 0.40 \\ 0.30 & 0.40 & 0.20 & 0.14 \\ \hline 0.30 & 0.20 & 0.10 & 0.16 \\ 0.20 & 0.25 & 0.40 & 0.30 \end{array}\right],$$

$$\underline{A}_2 = \left[\begin{array}{cc|cc} 0.11 & 0.02 & 0.22 & 0.01 \\ 0.03 & 0.03 & 0.01 & 0.09 \\ \hline 0.04 & 0.11 & 0.03 & 0.12 \\ 0.03 & 0.01 & 0.03 & 0.03 \end{array}\right],$$

$$\overline{A}_2 = \left[\begin{array}{cc|cc} 0.40 & 0.30 & 0.60 & 0.30 \\ 0.20 & 0.20 & 0.10 & 0.22 \\ \hline 0.25 & 0.40 & 0.20 & 0.36 \\ 0.15 & 0.10 & 0.10 & 0.12 \end{array}\right],$$

$$\underline{A}_3 = \left[\begin{array}{cc|cc} 0.09 & 0.09 & 0.02 & 0.31 \\ 0.03 & 0.08 & 0.09 & 0.08 \\ \hline 0.06 & 0.12 & 0.18 & 0.04 \\ 0.13 & 0.04 & 0.07 & 0.09 \end{array}\right],$$

$$\overline{A}_3 = \left[\begin{array}{cc|cc} 0.32 & 0.18 & 0.30 & 0.52 \\ 0.22 & 0.24 & 0.17 & 0.22 \\ \hline 0.15 & 0.31 & 0.32 & 0.13 \\ 0.33 & 0.27 & 0.21 & 0.13 \end{array}\right],$$

$$\underline{x}_0 = \begin{bmatrix} 1 \\ 3 \\ 3 \\ 2 \end{bmatrix}, \quad \overline{x}_0 = \begin{bmatrix} 8 \\ 6 \\ 11 \\ 7 \end{bmatrix}, \quad C = \left[\begin{array}{cc|cc} 1 & 0 & 0 & 0 \\ 0 & 1 & 0 & 0 \end{array}\right],$$

For a computer simulation, we consider the following subsystem matrices:

$$A_1 = \left[\begin{array}{cc|cc} 0.05 & 0.13 & 0.18 & 0.34 \\ 0.27 & 0.13 & 0.12 & 0.10 \\ \hline 0.15 & 0.04 & 0.08 & 0.12 \\ 0.11 & 0.17 & 0.34 & 0.22 \end{array}\right],$$

$$A_2 = \left[\begin{array}{cc|cc} 0.23 & 0.19 & 0.43 & 0.22 \\ 0.04 & 0.16 & 0.05 & 0.10 \\ \hline 0.20 & 0.14 & 0.18 & 0.14 \\ 0.07 & 0.02 & 0.09 & 0.04 \end{array}\right],$$

$$A_3 = \left[\begin{array}{cc|cc} 0.27 & 0.15 & 0.11 & 0.47 \\ 0.21 & 0.10 & 0.14 & 0.11 \\ \hline 0.12 & 0.14 & 0.29 & 0.11 \\ 0.16 & 0.19 & 0.15 & 0.13 \end{array}\right],$$

and the initial state

$$x_0 = \begin{bmatrix} 7.09 \\ 3.27 \\ 5.96 \\ 3.85 \end{bmatrix}.$$

Note that since $\underline{A}_i \geq 0$ ($i = 1, 2, 3$) and $\underline{x}_0 \geq 0$, the conditions stated in Assumption 2 are satisfied. Choosing the initial observer state $\omega_0^\ell$ and $\omega_0^u$ as

$$\omega_0^\ell = \begin{bmatrix} 2 \\ 1 \end{bmatrix}, \quad \omega_0^u = \begin{bmatrix} 12 \\ 8 \end{bmatrix},$$

the following interval reduced-order observer gain

$$L = \begin{bmatrix} 0.002 & 0.042 \\ 0.016 & 0.024 \end{bmatrix} \in \mathbf{R}_+^{2 \times 2}$$

satisfies assumptions (i)-(iv) of Theorem 2. Hence, it follows from Theorem 2 that there exists an interval reduced-order switched positive observer $\left(\hat{\Sigma}_\sigma^{dr\ell}, \hat{\Sigma}_\sigma^{dru}\right)$ with $\omega_0^\ell \leq \omega_0^u$ for $\Sigma_\sigma^d$ such that $\xi(k)(:= \hat{x}^u(k) - \hat{x}^\ell(k))$ is bounded and

$$0 \leq \hat{x}^\ell(k) \leq x(k) \leq \hat{x}^u(k) \ (k = 0, 1, \ldots)$$

under arbitrary switching, where

$$\hat{\Sigma}_\sigma^{dr\ell}: \begin{cases} \omega^\ell(k+1) = \hat{\underline{A}}_{\sigma(k)} \omega^\ell(k) + \underline{G}_{\sigma(k)} y(k), \\ \qquad\qquad\qquad\qquad\qquad \omega^\ell(0) = \omega_0^\ell \ (\geq 0), \\ \hat{x}^\ell(k) = \hat{C} \omega^\ell(k) + \hat{D} y(k), \end{cases}$$

and

$$\hat{\Sigma}_\sigma^{dru} : \begin{cases} \omega^u(k+1) = \hat{\overline{A}}_{\sigma(k)} \omega^u(k) + \overline{G}_{\sigma(k)} y(k), \\ \qquad\qquad\qquad\qquad \omega^u(0) = \omega_0^u \ (\geq 0), \\ \hat{x}^u(k) = \hat{C} \omega^u(k) + \hat{D} y(k). \end{cases}$$

Moreover, using the information from the proof of Theorem 1, the following uncertain interval reduced-order observers can be derived:

$$\underline{\hat{\Sigma}}_\sigma^{dr} : \begin{cases} \underline{\omega}(k+1) = \hat{A}_{\sigma(k)} \underline{\omega}(k) + G_{\sigma(k)} y(k), \\ \qquad\qquad\qquad\qquad \underline{\omega}(0) = \omega_0^\ell \ (\geq 0), \\ \underline{\hat{x}}(k) := \hat{C} \underline{\omega}(k) + \hat{D} y(k), \end{cases}$$

$$\overline{\hat{\Sigma}}_\sigma^{dr} : \begin{cases} \overline{\omega}(k+1) = \hat{A}_{\sigma(k)} \overline{\omega}(k) + G_{\sigma(k)} y(k), \\ \qquad\qquad\qquad\qquad \overline{\omega}(0) = \omega_0^u \ (\geq 0), \\ \overline{\hat{x}}(k) := \hat{C} \overline{\omega}(k) + \hat{D} y(k). \end{cases}$$

The results of Matlab simulations are illustrated in Figs. 3 and 4. Figure 1 shows the state trajectory and the estimate obtained from the designed observer, whereas Fig. 4 depicts the switching signal. We can see from Fig. 3 that the state $x(k)$ remains between the estimates $\hat{x}^\ell(k)$ and $\hat{x}^u(k)$,

$$0 \leq \hat{x}^\ell(k) \leq x(k) \leq \hat{x}^u(k) \ (k = 0, 1, \ldots),$$

under arbitrary switching. According to the chosen form of the output matrix of $\mathrm{rank} C = p = 2$, the state variables $x_1(k)$ and $x_2(k)$ match the first and the second element of the lower bound state estimate $\hat{x}^\ell(k)$ and the upper bound state estimate $\hat{x}^u(k)$, respectively. The other $(4-p) = 2$ components of the state vector are provided by the interval reduced-order observer with the estimation error quickly vanishing.

## 5. CONCLUDING REMARKS

In this paper, following a formal mathematical argument, interval reduced-order switched positive observers, constructed as $(n-p)$ dimensional positive lower and positive upper observers under arbitrary switching, for both continuous- and discrete-time uncertain switched positive linear systems were designed. It was also shown that the obtained results directly translate to both continuous- and discrete-time uncertain positive linear systems without switching, respectively. The observer properties, in particular robustness to interval-type uncertainty on the initial condition and system parameters, were illustrated in step-by-step design examples and numerical experiments conducted in both continuous- and discrete-time domain.

The observers provide state estimates under arbitrary switching. An interesting subject for further study would be to consider a state-dependent switching rule. Also, investigating effective algorithms to compute the observer gain $L \in \mathbf{R}_+^{(n-p) \times p}$ satisfying conditions (i)-(iv) of Theorems 1 and 2, would be a valuable future research direction. The presented material concentrated on the analytical aspects of observer design and property analysis. As a next step, application in practical control problems in the area of positive systems, e.g., inventory management in goods distribution networks [1], and switched positive systems, e.g., multi-modal transport systems [41], will be explored.

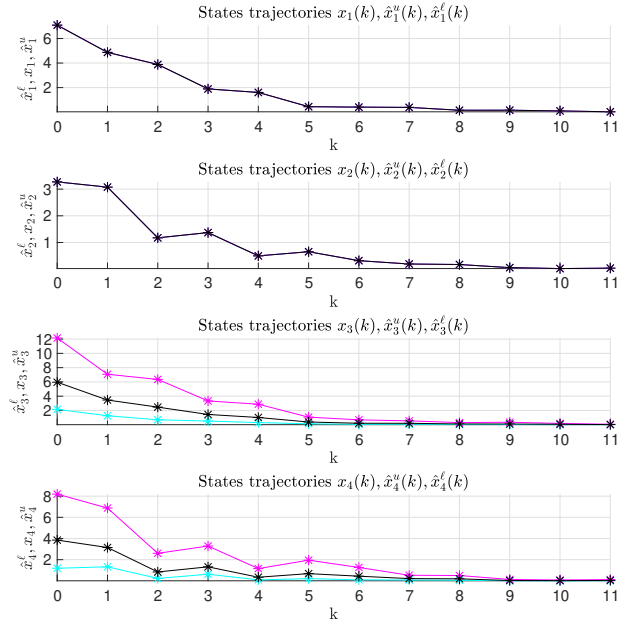

Fig. 3. System state $x(k)$ and estimates $\hat{x}^\ell(k)$, $\hat{x}^u(k)$ provided by the interval reduced-order switched positive observer – discrete-time case

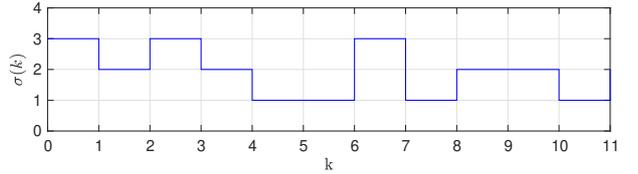

Fig. 4. Switching signal – discrete-time case

## ACKNOWLEDGMENT

The authors would like to express their gratitude to the Senior Editor, Associate Editor, and Reviewers for many helpful and valuable comments and suggestions. The work of N. Otsuka was supported in part by JSPS KAKENHI Grant Number 19K04443. The work of P. Ignaciuk was supported by a project "Robust control solutions for multi-channel networked flows" no. 2021/41/B/ST7/00108 financed by the National Science Centre, Poland.

## CONFLICT OF INTEREST

The authors declare that they have no conflict of interest.

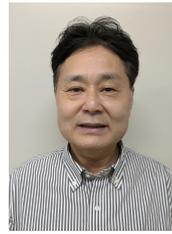

**Naohisa Otsuka** received the B.S. and M.S. degrees, both in Mathematical Sciences, from Tokyo Denki University (TDU) in 1984 and 1986, respectively. In 1992, he received the Doctor of Science in Mathematical Sciences from Graduate School of Science and Engineering of TDU. From 1986 to 1992 he served as a Research Associate in the Department of Information Sciences, TDU. In April 1992, he moved to the Institute of Information Sciences and Electronics, University of Tsukuba as a Research Associate and served as an Assistant Professor of the same University from 1993 to September 2000. From October 2000 to September 2003 he served as an Associate Professor in the Department of Information Sciences, TDU. From October 2003 to 2007 he served as a Professor in the same department. Since April 2007 he has been with the Division of Science in TDU as a Professor. His research interests include geometric control theory, switched systems, epidemic mathematical models, robust control, infinite-dimensional linear systems. He served as an associate editor of Journal of The Franklin Institute from 2005 to 2016. Dr. Otsuka is an editor of Mathematical Problems in Engineering and is a member of IFAC Technical Committee on Linear Control Systems. He is a member of the Society of Instrument and Control Engineers of Japan (SICE), IEEE and SIAM.

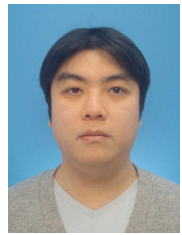

**Daiki Kakehi** received the B.S. degree in Science and Engineering and the M.S. degree from the Graduate School of Science and Engineering, both from Tokyo Denki University (TDU) in 2017 and 2019, respectively. He is currently working for a company in Japan. His research interests include switched linear systems, positive linear systems and observer design.

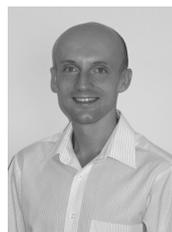

**PRZEMYSŁAW IGNACIUK** received the M.Sc. (with Hons.) degree in telecommunications and computer science and the Ph.D. (with Hons.) degree in control engineering and robotics from Lodz University of Technology, Łódź, Poland, in 2005 and 2008, respectively, and the Habilitation degree in computer science from Systems Research Institute Polish Academy of Sciences, Warsaw, Poland, in 2014. He worked for three years as a full-time Analyst and IT System Designer with the Telecommunications Industry. He then joined the Institute of Automatic Control, and since 2011, he has been with the Institute of Information Technology, Lodz University of Technology, where he is currently an Assistant Professor and the Head of a research group dedicated to complex system analysis and design. He has authored or coauthored one book, 15 monograph chapters, and over 180 journal and conference papers, in the area of control systems, communication and logistic networks. His research interests include networked control systems, robust control, dynamical optimization, and time-delay systems.